\documentstyle[preprint,aps,epsf]{revtex}

\begin{document}
\draft

\title{Amplitude Ratios and the Approach to Bulk Criticality in Parallel Plate Geometries}

\author{M. M. Leite}
\address{
Instituto de F\'{\i}sica Te\'orica, Universidade Estadual Paulista\\
01405-900 S\~ao Paulo, SP, Brazil
}
\author{M. Sardelich and M. D. Coutinho-Filho}
\address{ Laborat\'orio de F\'{\i}sica Te\'orica e Computacional,
Departamento de F\'{\i}sica,\\
Universidade Federal de Pernambuco 50670-901 Recife, PE, Brazil
}
\date{\today}
\maketitle
\begin{abstract}
We present analytical and numerical results for the specific heat and susceptibility amplitude ratios in parallel plate geometries. The results are derived using field-theoretic techniques suitable to describe the system in the bulk limit, i.e., $(L / \xi_{\pm} ) \gg 1$, where $L$ is the distance  between the plates and $\xi_{\pm}$ is the correlation length above (+) and below (-) the bulk critical temperature. Advantages and drawbacks of our method are discussed on the light of other approaches previously reported in the literature.\\
\end{abstract}

\large

\section{Introduction}
Since the advent of modern scaling concepts and renormalization-group
techniques the study of finite-size and surface effects on the behavior of
systems near or at criticality has attracted the attention of a number of
investigators$^{1}$. 

Fixing our interest in the case of a system confined between two infinite
($d-1$)-dimensional parallel plates distant $L$ from each other, we may
classify three well defined distinct regions in this problem. The first
one, where the scaling variable $(L / \xi_{\pm} ) \gg 1$, is characterized
by the dominance of bulk over surface and finite-size effects and the physics
is quasi d-dimensional. Here, $\xi_{\pm}$ specifies the critical correlation
length above (+) and below ($-$) the bulk critical temperature $T_{c}$.
The second region, where $(L / \xi_{\pm}) \ll 1$,
the system behaves as a quasi ($d-1$)-dimensional object. Finally, for $(L
/\xi_{\pm}) \sim 1$, the physics interpolates between a quasi
$d$- and a quasi ($d-1$)-dimensional system. A full description of the system
should therefore unveil very interesting crossover behaviors.

According to the region and phenomenon of interest, different field-theoretic
techniques have been devised to deal with such systems. Indeed, Diehl and
Dietrich$^{2,3}$ successfully implemented these techniques to study critical
and multicritical phenomena near surfaces within a finite momentum cutoff
regularization scheme. The use of dimensional regularization was shown$^{4,5}$ to
simplify the computational procedure and allowed the study of ordinary$^{2}$
and special transitions$^{3}$ through standard $\phi^{4}$-field theories under
Dirichlet (DBC) and Neumann (NBC) boundary conditions, respectively. The
former mimics very strong repulsive forces at the surface, thus preventing
order at it (a parameter $c$, which measures these forces$^{2}$, has fixed
point value $c^{\ast} = \infty$), whereas under the later boundary condition
both the surface and the bulk go critical simultaneously. The special
transition is in fact a multicritical point$^{3}$, $c^{\ast} = 0$, where the two
lines describing systems with repulsive ($c > 0$) and attractive forces ($c <
0$) at the surface, meet. In the later case, namely the extraordinary
transition$^{5,6}$, the surface undergoes a second order transition before
criticality sets in the bulk. Moreover, it has been shown$^{7}$ that a
scaling description holds so that the critical exponents associated with
excess surface singularities may be expressed completely in terms of bulk
exponents. However, it has also been shown$^{7}$ that fluctuations may
induce divergences at the surface and in these cases local quantities and associated
exponents must be defined resulting in new scaling relations. Therefore, in
treating these quantities at the multicritical point, Neumann boundary
conditions are valid only at the mean-field level$^{7}$.

On the other hand, in order to properly describe finite size effects using
field-theoretic techniques in critical systems subject, for example, to
periodic boundary conditions (PBC), Brezin and Zinn-Justin$^{8}$ and,
independently, Rudnick et al$^{9}$, introduced a method in which the
zero-momentum component is isolated whereas the other non-zero modes are
treated perturbatively. This methods has been largely used$^{10,11,12}$ and
generalized to study different boundary conditions. More recently$^{13}$,
some difficulties to treat critical systems below $T_{c}$ using this
technique have been circumvented.

Both finite size and surface effects are simultaneously present, except in
special circumstances such as for PBC where surface effects do not
contribute. At $T=T_{c}$, where Casimir forces manifest, these contributions
compete in a very special way, and powerful tools and methods such as
conformal invariance$^{14}$ and elaborated perturbation techniques$^{15}$
have been used to study this regime and the approach to
$T_{c}^{12,13,15,16}$. 

In this work we shall calculate specific heat and susceptibility amplitude
ratios using field-theoretic and $\in$-expansion methods$^{16}$ particularly suitable to
describe systems in the first regime mentioned above in which bulk behavior
dominates over surface and finite size contributions. The reported results
complement previous studies$^{16}$ and shed some light on the approach to
bulk criticality as the distance $L$ between the plates increases. 

For a system of volume $V = AL$, where $A$ is a ($d-1$)-dimensional surface
(layered geometry), the following asymptotic scaling form for the singular
part of the free energy density holds$^{17}$:
\begin{equation}
f( |t|, L) \sim \frac{1}{AL} Y (L | t|^{\nu} /a) =
y_{b} |t|^{d \nu} + y_{s} \frac{|t|^{(d-1) \nu}}{L} + \delta f (L |t|^{\nu} /
a) \hspace{0.5cm} ,
\end{equation}

\noindent
where $t = (T-T_{c})/T_{c}$, $\nu$ is the bulk correlation length exponent
and $a$ is the only non-universal metric factor. Using the hyperscaling
relation $d \nu = 2- \alpha$, one identifies the first term (proportional to
$y_{b}$) as the bulk contribution, the second one (proportional to $y_{s}$)
as the excess surface term and the last one as a finite-size correction term.
In the limit $( \xi_{\pm} /L) \ll 1$ one expects exponentially small
corrections from $\delta f$, whereas for $( \xi_{\pm} / L) \gg 1$
it compensates the bulk and surface contributions and  gives
rise to the Casimir effect$^{13-15}$ at $T = T_{c}$. 

From the above scaling assumptions the specific heat and susceptibility
should behave as
\begin{eqnarray} 
C(t,L) \sim |t|^{- \alpha} \; A_{\pm} (L|t|^{\nu}/a)
\hspace{0.5cm} ,\\
\chi (t,L) \sim |t|^{- \gamma} \; C_{\pm} (L|t|^{\nu}/a)
 \hspace{0.5cm} ,
\end{eqnarray}

\noindent
where $\alpha$ and $\gamma$ are the bulk critical exponents, but
even in the regime $L|t|^{\nu} >> 1$ one expects that excess surface and finite-size
contributions modify their critical amplitudes in a non-trivial manner.
In fact, the ratio of these amplitudes are quite sensitive in
identifying the universality class of a critical system, particularly in
numerical simulations$^{18}$ where one has to control both corrections to
scaling and surface and finite size effects.

In  Section II we explain our method and derive both the renormalized
free energy and the equation of state from which the above quantities can be
calculated. Finally, in Section III a discussion of the results and
conclusions are presented.

\section{Specific Heat and Susceptibility Critical Amplitudes}
In this section we shall use field-theoretic and renormalization-group
techniques to calculate the amplitude ratios of $C$ and $\chi$ in layered
geometries. We shall keep close contact with standard bulk $\phi^{4}-$field
theory$^{19}$ and whenever necessary to deal with the finite size of the
system we employ methods$^{16}$ which are particularly suitable in the regime
$(L / \xi_{\pm} ) \gg 1$.

\subsection{Renormalized Free Energy and Boundary Conditions}
We start by writing the expression for the one-loop renormalized Helmholtz
free energy density at the fixed point associated with the bulk critical
behavior of the system: 
\begin{eqnarray}
& & F(t,M;L) = \frac{1}{2} tM^{2} + \frac{1}{4!} u^{\ast} M^{4} + \frac{1}{4}
\left( t^{2} +  u^{\ast} t M^{2} + \frac{1}{4} u^{\ast^{2}} M^{4} \right) 
I_{sp} \nonumber \\
& & + \frac{1}{2L} \sum_{j} \int d^{d-1} q 
\ln \left[ 1 + (1/2) u^{\ast} M^{2} / (q^{2}+ \kappa^{2}_{j} + t) \right] 
\hspace{0.5cm}. 
\end{eqnarray}

\noindent
In the equation above $t, M \left( t_{0} = {\cal Z}_{\phi^{2}} t, \; M_{0} =
{\cal Z}^{1/2}_{\phi} M \right)$ are the renormalized (bare) reduced
temperature and order parameter, respectively, ${\cal Z}_{\phi^{2}}$,
${\cal Z}_{\phi}$ are renormalization functions, $u^{\ast}$ is the 
dimensionless renormalized coupling constant of a continuous (bulk) $\phi^{4}$
theory at the fixed point, $\vec{q}$ is a $(d-1)$-dimensional wave vector
along the direction parallel to the plate, $\kappa_{j} = \pi j /L$ are the
eigenvalues of the kinetic energy operator satisfying proper boundary
conditions (see below) and $I_{sp} = \in^{-1} [1+(\in /2)] + {\cal O} ( \in )$ is
the one-loop integral of a bulk $\phi^{4}$ theory evaluated at the symmetry
point using dimensional regularization, where $\in = 4 - d$. Notice that
taking the $\lim L \rightarrow \infty$ in Eq.(4) one obtains the standard
expression for the d-dimensional $\phi^{4}$ one-loop renormalized free
energy. 

In deriving $F(t,M;L)$ we have considered that the local field of a $\phi^{4}$ theory may satisfy periodic (PBC), Neumann (NBC) or Dirichlet (DBC) boundary conditions,
defined by: $ \phi ( \vec{\rho},z) = \phi ( \vec{\rho}, z+L)$, $(\partial
/ \partial z |_{z=0}) \phi ( \vec{\rho} , z) = ( \partial / \partial z |_{z=L})
\phi (\vec{\rho} , z)=0$ and $\phi ( \vec{\rho},z=0) = \phi ( \vec{\rho} , z=L)=0$, respectively, where $\vec{\rho}$ is a $(d-1)$-dimensional position vector perpendicular to the $z(dth)$ direction. It then follows that the sum in Eq.(4) have values $j=0$, $\pm 1$, $\pm 2$,..., for PBC, $j=0,1,2,...,$ for NBC, and $j=1,2,...,$ for DBC, respectively. The local field is Fourier transformed in the form$^{16}$ 
\begin{equation}
\phi ( \vec{\rho} , z) = \sum_{j} (2 \pi )^{1-d} \int d^{d-1}q \; exp (i
\vec{q}. \vec{\rho} ) \phi_{j} (\vec{q}) u_{j}(z) \hspace{0.5cm} ,
\end{equation}

\noindent
where $\phi_{j}(\vec{q})$ are plane waves parallel to the plate and
$u_{j}(z)$ are eigenfunctions of the kinetic energy operator
$(-d^{2}/dz^{2})$ with eigenvalues $\kappa^{2}_{j}$.  The bare order
parameter $M_{0}$ is thus the expectation value of the local field above. We call attention that the usual counterterms of a bulk $\phi^{4}$ theory are used to renormalize the free energy and that the boundary conditions are implemented
on the bare vertex functions. Details of the Feynman rules involving propagators and vertices can be found in Ref.(16).

\subsection{Specific Heat Amplitude Ratio}
Since the vertex function $\Gamma^{(0,2)}$ is additively renormalized, the
critical behavior (singular part) of the specific heat is calculated using
the expression$^{19}$:
\begin{equation}
C = A | t |^{- \alpha} = - \frac{\nu}{\alpha} B(u^{\ast})
- \Gamma^{(0,2)}_{R} \hspace{0.5cm} ,
\end{equation}

\noindent
where $B(u^{\ast})$ is the inhomogeneous
term of the renormalization group equation for $\Gamma^{(0,2)}_{R}$ and
\begin{equation}
\Gamma^{(0,2)}_{R} = \frac{\partial^{2}}{\partial t^{2}} F(t,M; L)
\hspace{0.5cm} .
\end{equation}

For $T>T_{c}$, $M=0$, and we find, using Eqs.(4) and (7) 
\begin{equation}
\Gamma^{(0,2)}_{R} (T>T_{c}) = - \frac{1}{2L} \sum_{j} \int \;
\frac{d^{d-1}q}{\left( q^{2} + \kappa^{2}_{j} + t \right)^{2}} + \frac{1}{2}
I_{sp} \hspace{0.5cm} ,
\end{equation}

\noindent
whereas for $T<T_{c}$ we use the value of $M$ at the coexistence curve,
namely $u^{\ast}M^{2}=-6t$, to obtain
\begin{equation}
\Gamma^{(0,2)}_{R} (T<T_{c}) = - \frac{3}{u^{\ast}} - \frac{2}{L}
\sum_{j} \int \; \frac{d^{d-1}q}{\left( q^{2} + \kappa^{2}_{j} + 2|t|
\right)^{2}} + 2 I_{sp} \hspace{0.5cm} .
\end{equation}

The one-loop integrals are evaluated using dimensional regularization 
and some useful formulae$^{16,20}$ to sum infinite series. We thus obtain for
the  boundary conditions of interest:
\begin{eqnarray}
& & \frac{1}{L} \sum_{j} \int \frac{d^{d-1}{q}}{\left( q^{2} + \kappa^{2}_{j} +
\tilde{t} \right)^{2}} = \tilde{t}^{- \in /2} \frac{1}{\in} \left( 1 -
\frac{\in}{2} \right) \nonumber \\
& & + 2 \pi^{-1/2} \left( \frac{2 \pi}{L} \right)^{d-4} \Gamma (d/2) \Gamma
[(5-d)/2)] sin [\pi (5-d)/2] f_{(5-d)/2} \left[ \frac{L
\tilde{t}^{(5-d)/2}}{2^{\sigma} \pi} \right] \nonumber \\
& & + \tau \frac{\pi^{1/2}}{2} \Gamma (d/2) \Gamma [(5-d)/2] \; \left[
\tilde{t}^{(d-5)/2}/L \right] \hspace{0.5cm} ,
\end{eqnarray}

\noindent
where $S_{d}/(2 \pi )^{d} \equiv 1$, $S_{d}$ being the area of
the d-dimensional unity sphere, $\tilde{t} = t + (1/2) u^{\ast}M^{2} = t ( \tilde{t} = 2|t|)$ for
$T>T_{c}$ ($T<T_{c}$), $\sigma =1$ for PBC, $\sigma =0$ for both NBC and DBC,
$\tau = 0, +1, -1$ for PBC, NBC and DBC, respectively, and
\begin{equation}
f_{\alpha} (a) = \int^{\infty}_{a} \frac{\left( u^{2}-a^{2} \right)^{-
\alpha} du}{exp (2 \pi u )-1} \hspace{0.5cm} , \hspace{0.5cm} \alpha < 1
\hspace{0.5cm} .
\end{equation}

Now using the $\in$-expansion$^{19}$ for the non-singular part of the specific heat,
$- ( \nu / \alpha ) B( u^{\ast} ) = (3/2 \in ) + (295/108)+ {\cal O}( \in)$,
we find the amplitudes above and below $T_{c}$:
\begin{eqnarray}
& & A_{+} = \frac{3}{\in 2} \left[ 1 + \in \frac{47}{54} + \in \frac{2^{2 -
\sigma}}{3} f_{1/2} \left( \frac{L t^{1/2}}{2^{\sigma} \pi} \right) +
\in \frac{\tau}{3} \frac{\pi}{Lt^{1/2}} \right] + {\cal O}(\in^{2}) \hspace{0.5cm} , \\
& & A_{-} = \frac{6}{\in 2^{\alpha}} \left[ 1 - \in \frac{7}{54} + 
\in \frac{2^{2 -\sigma}}{3} f_{1/2} \left( \frac{\sqrt{2} L
|t|^{1/2}}{2^{\sigma} \pi} \right) + \in \frac{\tau \pi}{3 \sqrt{2}L |t|^{1/2}} 
\right] + {\cal O}(\in^{2}) \hspace{0.5cm} , \nonumber \\  
\end{eqnarray}

\noindent
where $\alpha = \in /6 + {\cal O}(\in^{2})$.

From the above equations we finally obtain the specific heat amplitude ratio:
\begin{equation}
\frac{A_{+}}{A_{-}} = \frac{2^{\alpha}}{4} [1+ \in f (x) + \in S_{A} (x)] + {\cal O}(
\in^{2} )  \hspace{0.5cm} ,
\end{equation}

\noindent
where $f(x)$ and $S_{A}(x)$ are given by
\begin{eqnarray}
& & f(x) = \frac{2^{2- \sigma}}{3} \left[ f_{1/2} (x/2^{\sigma} \pi ) - f_{1/2} (
\sqrt{2} x / 2^{\sigma} \pi ) \right]  \hspace{0.5cm} ,\\
& & S_{A}(x) = \tau \frac{\pi}{3x} \left( 1 - \frac{1}{\sqrt{2}} \right)
\hspace{0.5cm} ,
\end{eqnarray}

\noindent
and $x = L/ \xi$, with $\xi = |t|^{-1/2}$. For $x \rightarrow \infty$, we can use the asymptotic limit$^{16,20}$ for $f_{1/2}(a)$ in Eq.(11) and write $f(x)$ in the more simplified form
\begin{equation}
f(x) = \frac{2^{1- \sigma} (2 \pi )^{1/2}}{3} \;
\left[ \frac{\exp (-2^{1- \sigma}x)}{(2^{1- \sigma} x)^{1/2}} -
\frac{\exp (-\sqrt{2}\hspace{2mm}2^{1- \sigma}x)}{(\sqrt{2}\hspace{2mm}2^{1- \sigma} x)^{1/2}}
\right] \; , \; x \rightarrow \infty \hspace{0.5cm} .
\end{equation}
\subsection{Susceptibility Amplitude Ratio}
Using Eq.(1) we obtain the following renormalized equation of state:
\begin{eqnarray}
H_{R} & = & \frac{\partial F}{\partial M} = tM + \frac{1}{6} u^{\ast} 
M^{3} + \frac{1}{2} u^{\ast} M \left( t + \frac{1}{2} u^{\ast} M^{2} \right)
\nonumber \\
& \times & \left[ I_{sp} - \sum_{j} \frac{1}{L} \int \frac{d^{d-1}q}{\left(
q^{2}+ \kappa^{2}_{j} \right) \left( q^{2} + \kappa^{2}_{j} + t + \frac{1}{2}
u^{\ast} M^{2} \right)} \right] \hspace{0.5cm} .
\end{eqnarray}

The one-loop integral is calculated similarly as for the specific heat:
\begin{eqnarray}
& & \frac{1}{L} \sum_{j} \int \frac{d^{d-1}{q}}{\left( q^{2}+ \kappa^{2}_{j}
\right) \left( q^{2} + \kappa^{2}_{j} + \tilde{t} \right)} =
\frac{1}{2} \Gamma [ 2 - ( \in /2 )] \Gamma ( \in / 2 ) \int^{1}_{0} dx
( \tilde{t} x)^{- \in /2} \nonumber \\
& & + 2 \pi^{-1/2} \left( \frac{2^{\sigma} \pi}{L} \right)^{d-4} \Gamma (d/2)
\Gamma [(5-d)/2]sin [ \pi (5-d)/2] \nonumber \\
& & \times \int^{1}_{0} dx f_{(5-d)/2} \left[\frac{L ( \tilde{t} x)^{(5-d)/2}}{2^{\sigma} \pi} \right] \nonumber \\
& & + \tau \frac{\pi^{1/2}}{2} \Gamma (d/2) \Gamma [(5-d)/2] \int^{1}_{0} dx
\left[ ( \tilde{t} x)^{(d-5)/2}/L \right] \hspace{0.5cm} .
\end{eqnarray}

By noticing that the first term in the right-hand side of Eq.(19) may be
written as $[ \in^{-1} - (1/2) \ell n \tilde{t} \; ]$, and using the $\in$-expansion
representation for $I_{sp}$, we obtain, to first order in $\in$, 
\begin{eqnarray}
H_{R} & = & t M + \frac{1}{6} u^{\ast} M^{3} + \frac{1}{4} u^{\ast} M
\tilde{t} \left\{ (1+ \ell n \tilde{t}) \right. \nonumber \\
& - & \left. 2 \int^{1}_{0} dy f_{1/2} \left[ \frac{L (y \tilde{t})^{1/2}}{2^{\sigma}
\pi} \right] + \tau \frac{\pi \tilde{t}^{-1/2} }{L} \right\} \hspace{0.5cm} , 
\end{eqnarray}

\noindent
where $u^{\ast} = (2/3) \in + {\cal O}( \in^{2})$ and $\tilde{t} = t + (1/2)
u^{\ast} M^{2}$.

The susceptibility amplitudes are then readily calculated from
\begin{equation}
\chi^{-1} = \left( C |t|^{- \gamma} \right)^{-1} = \Gamma^{(2,0)}_{R} =
\frac{\partial H_{R}}{\partial M} \hspace{0.5cm} .
\end{equation}

\noindent
As before, for $T > T_{c}$, $M=0$, and for $T<T_{c}$ we use
$u^{\ast}M^{2}=- 6t$. The amplitudes above and below $T_{c}$ thus read:
\begin{eqnarray}
& & C_{+} = 1 - \frac{\in}{6} + \frac{\in}{3} \int^{1}_{0} dy f_{1/2} \left(
\frac{xy^{1/2}}{2^{\sigma} \pi} \right) + \in \frac{\tau \pi}{6x} +
{\cal O} (\in^{2}) \hspace{0.5cm} ,  \\
& & C_{-} = \frac{1}{2} \left\{ 1 - \frac{\in}{6} (4+ \ell n 2) + 
\frac{\in}{3} \int^{1}_{0} dy f_{1/2} 
\left[ \frac{x(2y)^{1/2}}{2^{\sigma} \pi} \right] + 
\in \frac{\tau \pi}{6 \sqrt{2}x} \right\} + {\cal O}(\in^{2}) \; , \nonumber \\
\end{eqnarray}

\noindent
where $x = L/ \xi$. 

Using Eq.(11) and performing the integrations in $y$, we find the
susceptibility amplitude ratio:
\begin{equation}
\frac{C^{+}}{C_{-}} = 2^{\gamma -1} \frac{\gamma}{\beta} + \in h (x) + \in
S_{C}(x) + {\cal O}( \in^{2} ) \hspace{0.5cm} ,
\end{equation}

\noindent
where $\gamma = 1+ ( \in / 6) + {\cal O}( \in^{2})$, $\beta = 1/2 - (\in /6) + {\cal O}(\in^{2})$ and 
\begin{eqnarray}
h(x) & = & \frac{2^{\sigma +1} \pi}{3x} 
\left[ \frac{(1- \sqrt{2})}{24} - \int^{\infty}_{a} du  g(a/u) 
+ \sqrt{2} \int^{\infty}_{b} du g(b/a) \right]  , \\
S_{C}(x) & = & \tau \frac{\pi}{3x} \left( 1 - \frac{1}{\sqrt{2}} \right) \hspace{0.5cm} ,
\end{eqnarray}

\noindent
with $a = \sqrt{2}x/2^{\sigma} \pi$, $b = x/2^{\sigma} \pi$ and 
\begin{equation}
g(c/u) = \frac{u cos [arc \; sin (c/u)]}{e^{2 \pi u} -1} \hspace{0.5cm} .
\end{equation}

\section{Discussion and Conclusions}
First, we should point out that the main step in our approach is the
representation$^{16,20}$ used to evaluate the discrete sums in Eqs.(10) and
(19). It has proved very useful in different field-theoretic contexts$^{20}$
and here it clearly helps to split the bulk, surface and finite size
contributions, as required by scaling [see Eqs.(1-3)], in a rather simple way, though its range of effectivenes precludes direct access to the Casimir effect.

Second, our starting renormalized free energy, Eq.(4), does not consider any
distortion of the order parameter profile, i.e., our description is
restricted to calculating the effect of the boundary conditions on bulk
quantities as a result of fluctuations, i.e., in the amplitude ratios, Eqs. (14) and (24), excess surface and finite size contributions are of ${\cal O}(\in )$. Nevertheless, we observe that if the excess surface contributions for NBC in Eqs.(12) and (13) are isolated, we find $(A_{+}/A_{-})_{s} = 2^{-3/2} + {\cal O}(\in )$, which is the same result derived in Ref.(5) for the special transition (the excess surface specific heat exponent is $\alpha_{S} = \alpha + \nu$). This is so because in
this particular case there is no distortion of
the order parameter at the mean-field level. Notice also that, above $T_{c}$,
$(A_{+})_{sp}/(A_{+})_{ord} = - 1 + {\cal O} (\in^{2})$, where
$(A_{+})_{sp,ord}$ refer to the specific heat amplitudes at the special (NBC)
and ordinary (DBC) transitions, a result already derived in Ref.(4). As for the excess surface contributions for the susceptibility amplitudes we also notice that the last term in Eq.(20) is consistent with the result found in Ref.(7) for the special transition above $T_{c}$ $(\gamma_{S} = \gamma + \nu )$, since again no distortion of the order parameter is necessary in this case. 

From the discussion above and the derived results in Section II, particularly
Eqs.(14-17) and (24-27), it is clear that in the regime $L / \xi_{\pm} \gg
1$, and to first-order in an $\in$-expansion, the specific heat and
susceptibility display singularities well described by bulk exponents, but
with amplitudes sensitive to the boundary conditions which manifest as excess
surface and finite size contributions. Notice also that these fluctuation
effects result quite effectively from the difference between the amplitudes
of the correlation length above and below $T_{c}$, which satisfy 
$( \xi_{0,+} / \xi_{0,-} )= \sqrt{2} + {\cal O}(\in)$. 

In Figs. 1, 2 and 3 we plot the scaling functions $f(x)$ and $h(x)$, numerically evaluated in $d=3$, as defined by Eq.(11,15,25-27). They both decay very rapidly to zero as $x = L t^{1/2}$ increases, in agreement with the asymptotic result$^{16,20}$ for $f_{1/2}(x)$, $x \rightarrow \infty$. In fact, for PBC and $x=7$ we find $f_{p}(x) \simeq 2,7 \times 10^{-4}$ both by using the numerical estimate and the asymptotic result predicted in Eq.(17). For this value of $x$ it is indeed expected$^{18}$ that these corrections to the bulk limit are indeed
negligible. Notice that for NBC the magnitude of the scaling functions are
the same in our one-loop approximation, in agreement with Ref.(16). However,
a two-loop calculation shows$^{16}$ that they slightly differ if the same
regime of validity applies. However, as $x \rightarrow 0$ our approach does not correctly describe the Casimir effect, as shown in Figs. 1 and 2: $f(x)$
diverges and $h(x)$ approaches zero, whereas in a correct treatment$^{13-15}$
both tend to a constant value, the Casimir amplitudes for each case. This
failure for $x < 1$ has already been pointed out by Nemirovsky and Freed$^{16}$. Here, our results deal with issue in a more quantitative way, thus evidencing the advantages, limitations and drawbacks of the method.
 
For comparison, we also plot in Fig. 1 the excess surface contribution for the
cases of both NBC and DBC, i.e., $\left| S_{A}(x) \right| = \left| S_{C} (x)
\right| \equiv S(x)$. As clearly seen from Figs. 1 and 2, as $x \rightarrow \infty$ these contributions are the leading ones$^{11,15}$ modifying the amplitudes of both the bulk specific heat and susceptibility. Notice that they having the same magnitude and decay as $x^{-1}$ ( this is fortuitously true only because these effects are treated as fluctuation contributions to the bulk limit, see Eqs. (16,26) ).

Finally, in Fig. 4 we plot the difference between the excess surface
contribution and the scaling function for NBC. We see that this
difference  ``almost'' saturates as $x \rightarrow 0$, as expected in the Casimir effect, but lostly it diverges for very small values of $x$ (a $x^{-1}$ dependence is in fact expected from Eq. (11) as $x \rightarrow 0$, but an extra ln$x$ contribution precludes a good description of the Casimir effect).  

In summary, we have presented a field-theoretic description of the approach
to bulk criticality in parallel plate geometries, in which excess surface and
finite-size contributions appear as a result of fluctuations and depend on
the boundary condition imposed on the system. Despite the fact that other
more general methods to deal with finite systems do exist, our approach is
probably the simplest one in the regime \mbox{$L / \xi_{\pm} \gg 1$.}

\acknowledgments

We thank A. M. Nemirovsky for collaboration in an early stage of this work and several fruitful discussions. M. M. L. acknowledges FAPESP (State of S. Paulo Foundation) for financial support, under grant number 96/03546-7. M. D. C.-F. and M. S. acknowledge FINEP, CNPQ, CAPES and FACEPE (Brazilian agencies) for financial support. 
\newpage

\noindent
{\bf REFERENCES}

\begin{enumerate}
\item See, e.g., M.E. Fisher, in {\em Critical Phenomena}, edited by M.S.
Green (Academic, London, 1971), p.1; V. Privman in: {\em Finite Size Scaling
and Numerical Simulation of Statistical Systems}, edited by V. Privman (World
Scientific Singapore, 1990) p.1; M.N. Barber in: {\em Phase Transitions and
Critical Phenomena}, edited by C. Domb and J.L. Lebowitz (Academic, 1983),
vol.VIII, p.145; H.W. Diehl {\em ibid}, (1986) vol.X, p.75; S. Dietrich, {\em
ibid}, (1988) vol.XII, p.1.
\item H.W. Diehl and S. Dietrich, Z. Phys. B-Condensed Matter \underline{42},
65 (1981); {\em ibid} (E) \underline{43}, 281 (1981); {\em ibid}, Phys. Lett.
\underline{80} A, 408 (1980).
\item H.W. Diehl and S. Dietrich, Z. Phys. B-Condensed Matter \underline{50},
\underline{117} (1983).
\item Y.Y. Goldschmidt and D. Jasnow, Phys. Rev. B \underline{29}, 3990
(1984). \item E. Eisenriegler, J. Chem. Phys. \underline{81}, 4666 (1984); E.
Eisenriegler and P.J. Upton, {\em ibid}, (E) \underline{98}, 3582 (1993).
\item P. Upton, Phys. Rev. B \underline{45}, 8100 (1992); E. Eisenriegler and
M. Stapper, Phys. Rev. B \underline{50}, 10009 (1994).
\item H.W. Diehl, G. Gompper and W. Speth, Phys. Rev. B \underline{31}, 5841
(1985); H.W. Diehl and M. Shpot, Phys. Rev. Lett. \underline{73}, 3431
(1994). See also, H.W. Diehl, in Ref.(1).
\item E. Br\'{e}zin and J. Zinn-Justin, Nuclear Physics B257 $[$FS14$]$, 867
(1985). 
\item J. Rudnick, H. Guo and D. Jasnow, J. Stat. Phys. \underline{41}, 353
(1985). 
\item E. Eisenriegler and R. Tomaschitz, Phys. Rev. B \underline35{}, 4876
(1987); E. Eisenriegler, Z. Phys. B-Condensed Matter \underline{61}, 299
(1985). 
\item W. Huhn and V. Dohm, Phys. Rev. Lett. \underline{61}, 1368 (1988); V.
Dohm, Physica Scripta T \underline{49}, 46 (1993) and references therein.
\item For crossover effects, see: D. O'Connor and C.R. Stephens, Nucl. Phys.
B \underline{360}, 297 (1991); {\em ibid}, J. Phys. A: Math Gen.
\underline{25} , 101 (1992); {\em ibid}, Phys. Rev. Lett. \underline{72}, 506
(1994). 
\item X.S. Chen, V. Dohm and N. Schultka, Phys. Rev. Lett. \underline{77},
3641 (1996); A. Esser, V. Dohm and X.S. Chen, Physica A \underline{222}, 355
(1995) and references therein.
\item J.L. Cardy, Phys. Rev. Lett. \underline{65}, 1443 (1990); T.W.
Burkhardt and T. Xue, {\em ibid} \underline{66}, 895 (1991); E. Eisenriegler
and U. Ritschel, Phys. Rev. B \underline{51}, 13717 (1995); T.W. Burkhardt and
E. Eisenriegler, Phys. Rev. Lett. \underline{74}, 3189 (1995).
\item M. Krech and S. Dietrich, Phys. Rev. Lett. \underline{66}, 345 (1991);
{\em ibid}, (E) \underline{67}, 1055 (1991); {\em ibid}, Phys. Rev. A \underline{46}, 1886 (1992); E. Eisenriegler, M.
Krech and S. Dietrich, Phys. Rev. Lett. \underline{70}, 619 (1993); {\em
ibid}, (E) \underline{70}, 2051 (1993); M. Krech, Phys. Rev. E
\underline{56}, 1642 (1997); {\em ibid}, {\em The Casimir Effect in Critical
Systems} (World Scientific, Singapore, 1994), and references therein.
\item A.M. Nemirovsky and K.F. Freed, Nucl. Phys. B \underline{270} 
[FS16], 423 (1986); {\em ibid}, J. Phys. A \underline{18}, L319 (1985); M.M.
Leite, A.M. Nemirovsky and M.D. Coutinho-Filho, J. Magn. Magn. Mat.
\underline{104-107}, 181 (1992).
\item See, e.g. V. Privman and M.E. Fisher, Phys. Rev. B \underline{30}, 322
(1984); V. Privman, Phys. Rev. B \underline{38}, 9261 (1988).
\item M. Caselle and M. Hasenbusch, J. Phys. A: Math. Gen. \underline{30},
4963 (1997); X.S. Chen, V. Dohm and A.L. Talapov, Physica A \underline{232},
375 (1996).
\item See, e.g., D.J. Amit, {\em Field Theory, the Renormalization Group and
Critical Phenomena} (World Scientific, Singapore, 1984).
\item See, e.g., N.D. Birrell and L.H. Ford, Phys. Rev. D \underline{22}, 330
(1980); D.J. Toms, {\em ibid}, D \underline{21}, 928 (1980).
\end{enumerate}

\newpage
\section*{Figure Captions}
\noindent
{\bf Fig. 1.} Amplitude of the excess surface contribution (solid line), from Eqs. (16) and (26), and scaling functions $f_{p}$ (dashed line) and $f_{N,D}$ (dotted line) for periodic, Neumann and Dirichlet boundary conditions, respectively, numerically evaluated using Eqs. (15) and (11) in $d=3$, as functions of $x = L/ \xi$.\\

\noindent
{\bf Fig. 2.} Same as in Fig. 1 for $2\leq x \leq 7$.\\

\noindent
{\bf Fig. 3.} Scaling function $h_{p}$ (dashed line) and $h_{N,D}$ (dotted line) for periodic, Neumann and Dirichlet boundary conditions, respectively, numerically evaluated using Eqs. (25) and (27) in $d=3$, as function of $x = L/ \xi$.\\

\noindent
{\bf Fig. 4.} Difference (apart from a  minus sign) between the excess surface contribution and the scaling function $f_{N}$ for Neumann boundary condition as function of $x = L/ \xi$.

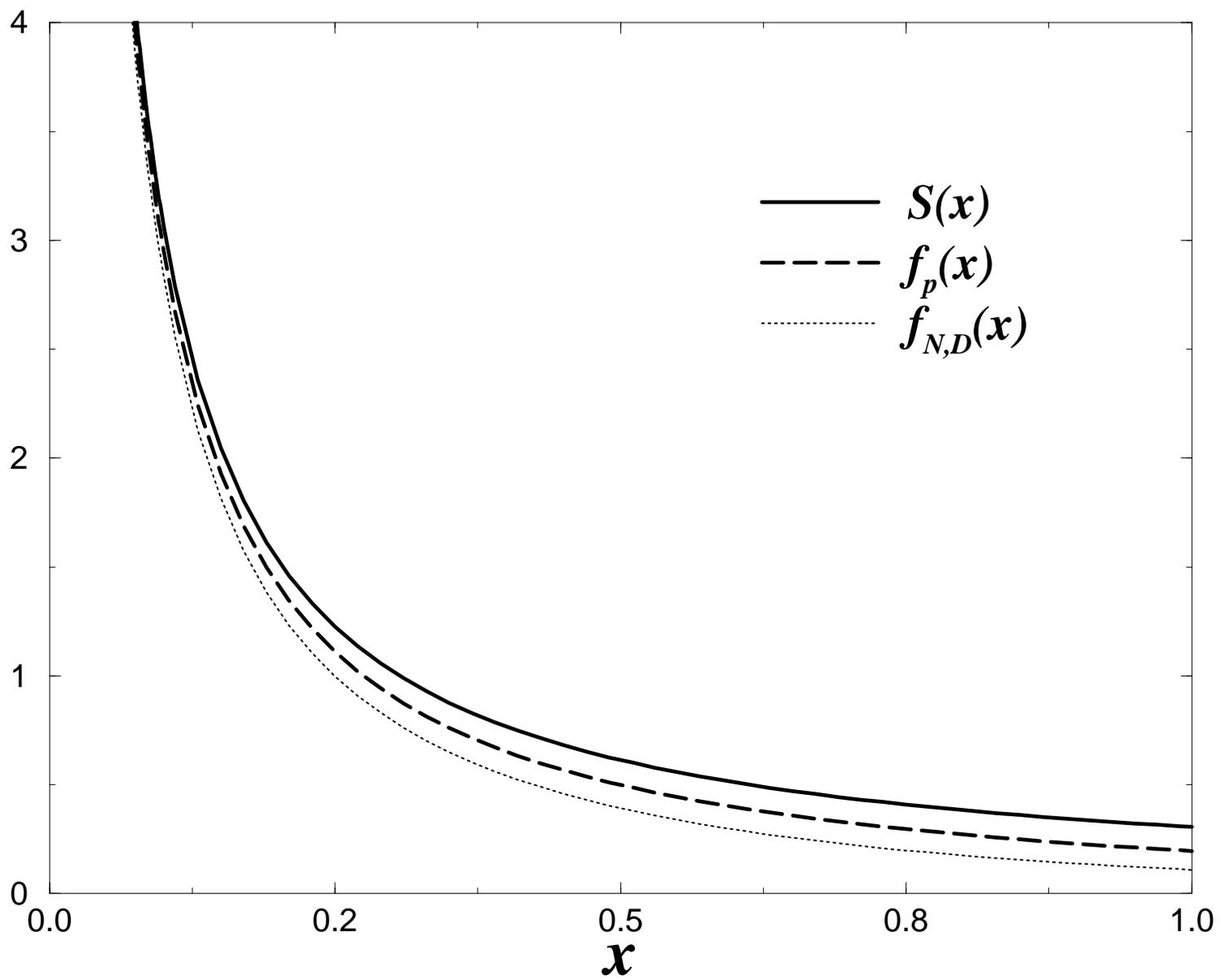
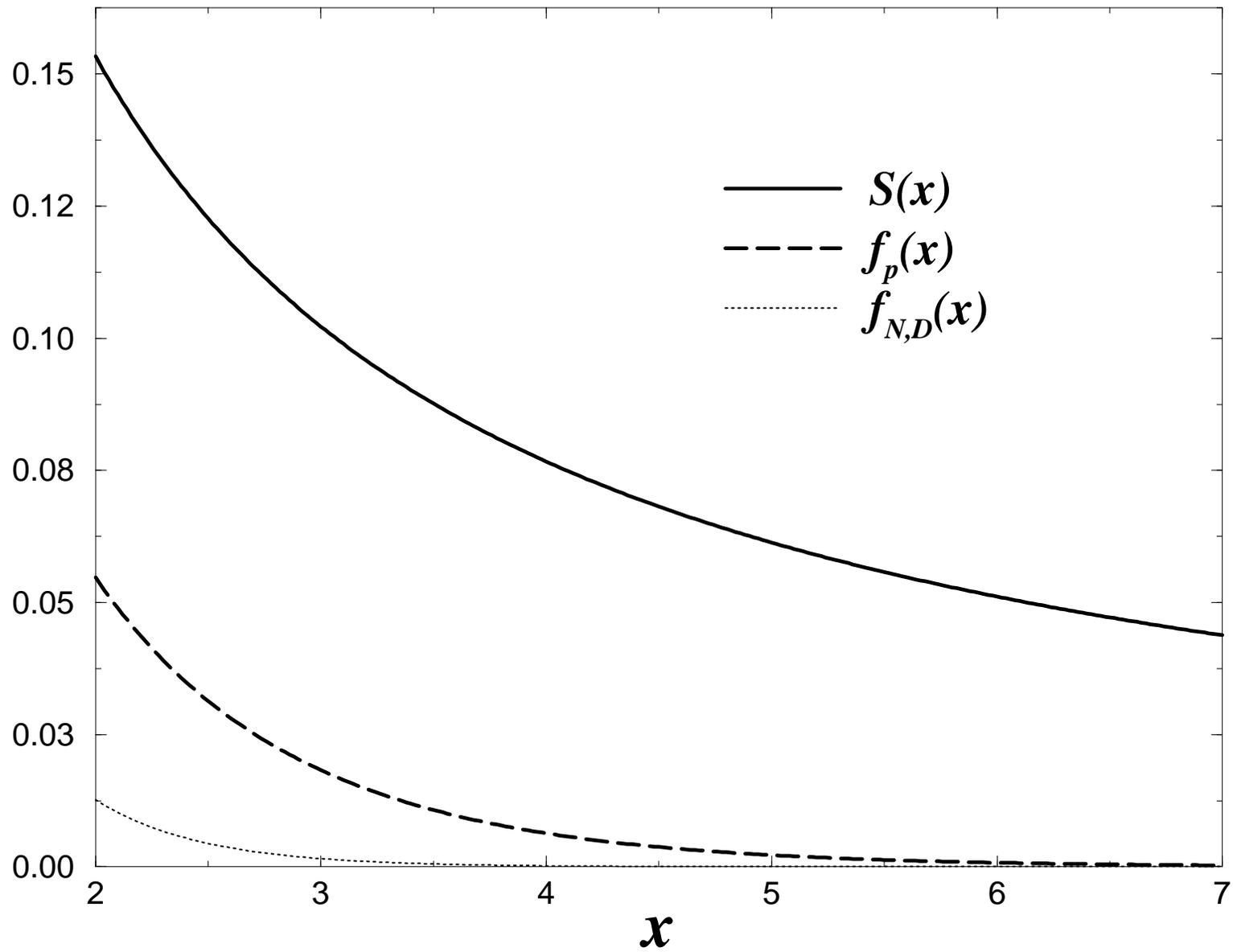
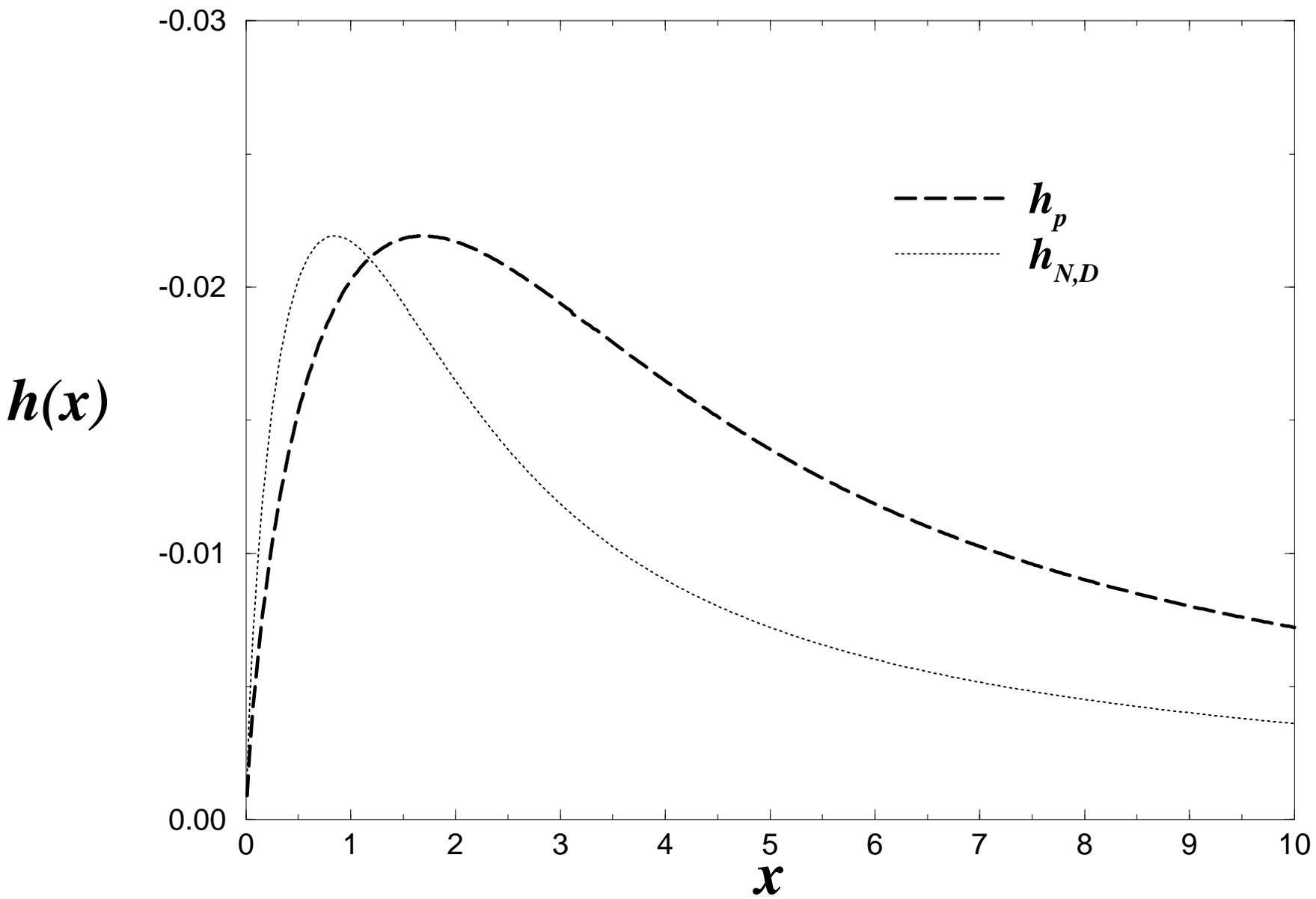
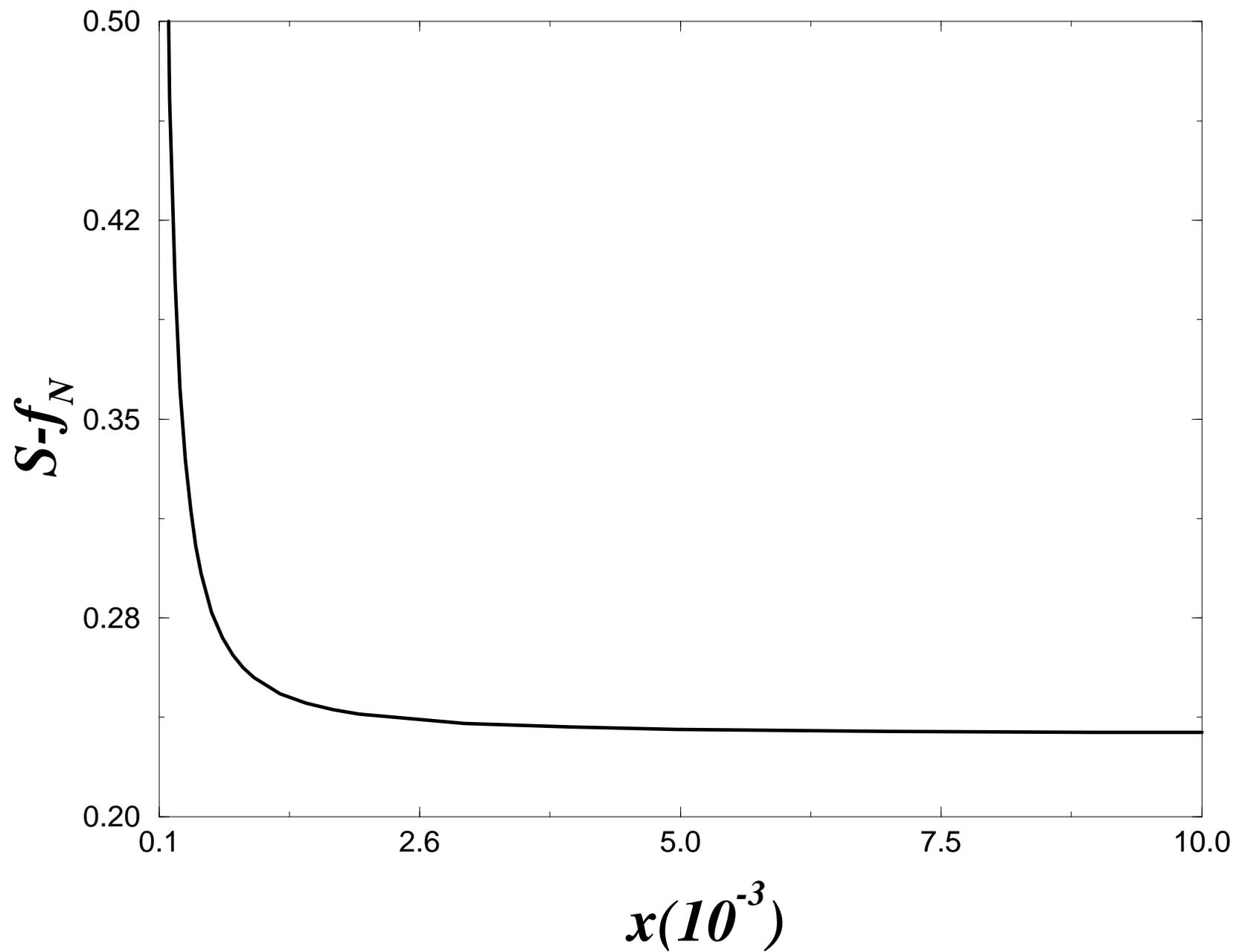
\end{document}